\documentclass[twocolumn]{aastex631}

\usepackage{placeins}
\usepackage{xspace}
\usepackage{natbib}
\bibpunct{(}{)}{;}{a}{}{,}
\usepackage{graphicx}
\usepackage{xcolor}
\usepackage{multirow}
\usepackage{amsmath}
\usepackage{lipsum}
\usepackage{wrapfig,lipsum,booktabs}

\usepackage{makecell}
\usepackage{float}
\usepackage{makecell}
\usepackage{placeins}

\usepackage[table]{xcolor}
\usepackage{booktabs}

\newcommand{\NH}{\ensuremath{N_{\mathrm{H}}}\xspace}

\newcommand{\slab}{{\tt slab}\xspace}

\newcommand{\amol}{{\tt amol}\xspace}
\newcommand{\hot}{{\tt hot}\xspace}

\newcommand{\xmm}{{XMM-\textit{Newton}}\xspace}
\newcommand{\chandra}{\textit{Chandra}\xspace}
\newcommand{\spex}{{\tt SPEX}\xspace}

\newcommand{\fei}{\ion{Fe}{1}\xspace}
\newcommand{\feii}{\ion{Fe}{2}\xspace}
\newcommand{\feiii}{\ion{Fe}{3}\xspace}

\newcommand{\siii}{\ion{Si}{2}\xspace}
\newcommand{\siiii}{\ion{Si}{3}\xspace}
\newcommand{\sii}{\ion{Si}{1}\xspace}
\newcommand{\sixiii}{\ion{Si}{13}\xspace}
\newcommand{\six}{\ion{Si}{10}\xspace}
\newcommand{\sixi}{\ion{Si}{11}\xspace}
\newcommand{\sixii}{\ion{Si}{12}\xspace}
\newcommand{\mgi}{\ion{Mg}{1}\xspace}
\newcommand{\mgii}{\ion{Mg}{2}\xspace}
\newcommand{\mgiii}{\ion{Mg}{3}\xspace}
\newcommand{\mgx}{\ion{Mg}{10}\xspace}
\newcommand{\mgxi}{\ion{Mg}{11}\xspace}
\newcommand{\mgix}{\ion{Mg}{9}\xspace}

\newcommand{\oi}{\ion{O}{1}\xspace}
\newcommand{\oii}{\ion{O}{2}\xspace}

\newcommand{\oiv}{\ion{O}{4}\xspace}
\newcommand{\ov}{\ion{O}{5}\xspace}
\newcommand{\ovi}{\ion{O}{6}\xspace}
\newcommand{\ovii}{\ion{O}{7}\xspace}
\newcommand{\oviii}{\ion{O}{8}\xspace}

\newcommand{\nei}{\ion{Ne}{1}\xspace}
\newcommand{\neii}{\ion{Ne}{2}\xspace}
\newcommand{\neiii}{\ion{Ne}{3}\xspace}

\newcommand{\nev}{\ion{Ne}{5}\xspace}
\newcommand{\nevi}{\ion{Ne}{6}\xspace}
\newcommand{\nevii}{\ion{Ne}{7}\xspace}
\newcommand{\neviii}{\ion{Ne}{8}\xspace}
\newcommand{\neix}{\ion{Ne}{9}\xspace}

\begin{document}

\title{Dust grain chemistry in the diffuse ISM towards the black hole transient GX 339-4}

\author[0000-0002-1049-3182]{I. Psaradaki}
\affiliation{MIT Kavli Institute for Astrophysics and Space Research, 70 Vassar Street, Cambridge, MA 02139}
\affiliation{European Space Agency (ESA), European Research and Technology Centre (ESTEC), Keplerlaan 1, 2201 AZ Noordwijk, The
Netherlands}
\affiliation{SRON Space Research Organisation Netherlands, Niels Bohrweg 4, 2333 CA Leiden, the Netherlands}

\author{L. Corrales}
\affiliation{LSA, University of Michigan, 1085 S University Ave, Ann Arbor, MI 48109}

\author{E. Costantini}
\affiliation{SRON Space Research Organisation Netherlands, Niels Bohrweg 4, 2333 CA Leiden, the Netherlands}
\affiliation{Anton Pannekoek Institute, University of Amsterdam, Postbus 94249, 1090 GE Amsterdam, The Netherlands}

\author{P. Draghis}
\affiliation{MIT Kavli Institute for Astrophysics and Space Research, 70 Vassar Street, Cambridge, MA 02139}

\author{J. A. Garc\'ia}
\affiliation{X-ray Astrophysics Laboratory, NASA Goddard Space Flight Center, Greenbelt, MD 20771, USA}
\affiliation{Cahill Center for Astrophysics, California Institute of Technology, Pasadena, CA 91125, USA}

\author{E. Gatuzz}
\affiliation{Max-Planck-Institut f\"ur extraterrestrische Physik, Gie{\ss}enbachstra{\ss}e 1, 85748 Garching, Germany}

\author{P. Kosec}
\affiliation{Center for Astrophysics — Harvard \& Smithsonian, Cambridge, MA, USA}

\author{G. Mastroserio}
\affiliation{Scuola Universitaria Superiore IUSS Pavia, Palazzo del Broletto, piazza della Vittoria 15, I-27100 Pavia, Italy}

\author{M. Mehdipour}
\affiliation{LSA, University of Michigan, 1085 S University Ave, Ann Arbor, MI 48109}
\affiliation{Space Telescope Science Institute, 3700 San Martin Dr, Baltimore, MD 21218, USA}

\author{F. Paerels}
\affiliation{Columbia Astrophysics Laboratory and Department of Astronomy, Columbia University, 550 West 120th St., New York, NY 10027, USA}

\author{D. Rogantini}
\affiliation{Department of Astronomy and Astrophysics, The University of Chicago, Chicago, IL 60637}

\author{N. Schulz}
\affiliation{MIT Kavli Institute for Astrophysics and Space Research, 70 Vassar Street, Cambridge, MA 02139}

\author{S. Zeegers}
\affiliation{SRON Space Research Organisation Netherlands, Niels Bohrweg 4, 2333 CA Leiden, the Netherlands}
\affiliation{Anton Pannekoek Institute, University of Amsterdam, Postbus 94249, 1090 GE Amsterdam, The Netherlands}

\begin{abstract}

We present results on X-ray absorption and the dust grain chemistry in the diffuse interstellar medium (ISM), based on a new Cycle 25 \textit{Chandra} High Energy Transmission Grating Spectrometer (HETGS) observational campaign targeting the black hole transient GX~339--4. The X-ray source offers an optimal combination of moderate hydrogen column density and high X-ray flux, enabling the first detailed simultaneous fitting of the photoabsorption edges of Fe, O, Si, and Mg which are key elemental constituents of interstellar dust. We performed a joint spectral analysis of \textit{Chandra}/HETGS data and archival observations from the Reflection Grating Spectrometer (RGS) on board XMM-Newton. We found that the dust grain chemical composition along this diffuse Galactic line of sight is best described by the silicate Mg-rich amorphous pyroxene ($\rm Mg_{0.75}Fe_{0.25}SiO_{3}$) and metallic iron. We also discuss the elemental abundances and depletions of Fe, O, Si, and Mg, and the presence of absorption features in the X-ray spectrum of this source associated with highly ionised plasma.

\end{abstract}

\section{Introduction}

Understanding the chemistry of the interstellar medium (ISM) is key to tracing the evolution of stars and galaxies. Though interstellar dust makes up only ~1–2\% of the ISM by mass, it plays a critical role across all stages of stellar evolution—from evolved stars and supernovae to protoplanetary disks. Understanding the composition, origin, and evolution of interstellar dust is a key question in modern astrophysics \cite[e.g.][]{Drainebook, bookWhittet}.

Cosmic dust grains vary in chemical composition, lattice structure, and physcial size. They are primarily composed of abundant elements such as C, O, Si, Mg, and Fe, while more refractory elements like Ti, Ca, and Al are present in smaller amounts and are highly depleted from the gas phase \citep{Jenkins2009}. Dust is broadly classified into carbonaceous and silicate types, but also includes oxides (e.g., $\mathrm{Fe_2O_3}$), sulfides (e.g., $\mathrm{FeS}$), carbides, and metallic iron \citep{Drainebook}. More recent work by \citet{Hensley_2023} proposes that dust grains can be treated as a composite of silicate and carbonaceous material, a framework known as the astrodust model.

Structurally, dust can be crystalline or amorphous. Crystalline grains have long-range atomic order, whereas amorphous grains exhibit a disordered atomic structure. Due to the harsh conditions of the ISM, most interstellar silicates are found in amorphous form—particularly olivines and Mg-rich pyroxenes—based on mid-IR and X-ray absorption studies \citep{Kemper2004, Min2007, Zeegers2017, Zeegers2019, Rogantini2019, Rogantini2020, Psaradaki2022}.

 Silicate dust is an important component of the life cycle of matter in the ISM. Silicates mainly consist of the abundant astrophysical elements Si, Mg, Fe and O, and they form in the winds of Asymptotic Giant Branch (AGB) stars, get modified, destroyed, and potentially reformed in the diffuse ISM. Silicates are also an important dust component of proto-planetary and debris disks, can regulate the thermal structure of the ISM and provide the surface for chemical reactions \citep{Henning2010}. 

Pyroxenes, $\rm (Mg,Fe)SiO_{3}$, and olivines, $\rm (Mg,Fe)_{2}SiO_{4}$, are two important groups of minerals that are expected to describe the chemical composition of interstellar dust \citep{Henning2010}. The exact ratio of olivines to pyroxenes in dust along with the Mg:Fe ratio in silicates is not yet fully constrained. This an important area of study as Mg-rich silicates (rather than Fe-rich) could give an explanation for the high amount of Mg found in cometary and circumstellar grains \citep{Min2007}.
Additional dust reservoirs beyond silicates, such as ferromagnetic inclusions (e.g., $\mathrm{Fe_3O_4}$) and iron sulfides, have been widely discussed in the literature as potential dust carriers.

A powerful way to study the ISM and the chemistry of interstellar dust is provided by the X-ray band \cite[e.g.][]{Schattenburg1986, Wilms, Schulz2002, Juett2004, Ueda2005, Lee2005}. 
The X-ray energy band includes the photoabsorption edges of key elements incorporated in dust grains, including Si, Mg, Fe, and O. High-resolution X-ray spectroscopy allows us to study the abundances and depletions of these elements along different density environments in our Galaxy. The X-ray spectra of X-ray binaries can be used as a background light to study the atomic and solid species of the ISM along different lines of sight. In particular, low-mass X-ray binaries (LMXB) are the ideal sources due to their high flux and smooth continuum, enabling observations of uncontaminated and well-defined dust absorption features.

X-ray absorption fine structures (XAFS) are spectroscopic features observed near the photoelectric absorption edges of solid material (dust), and their shape is the ultimate footprint of the dust chemical composition and lattice structure. In the literature there are several studies of XAFS \cite[e.g.][]{Ueda2005, Lee2009b, Costantini2012, Pinto2010, Pinto2013, Corrales2016, Zeegers2017, Zeegers2019, Rogantini2018, Rogantini2019, Psaradaki2020, Psaradaki2022, Costantini2022}. More recent studies, such as those of \cite{Rogantini2019} and \cite{Zeegers2019}, which employed updated dust extinction cross sections derived from laboratory data, indicate that amorphous olivine is the dominant dust species in the dense ISM. In addition, it has been found that the diffuse ISM is composed predominantly of amorphous pyroxene, and in smaller amount Fe metallic \citep{Psaradaki2022}.

So far, each work focused on modelling of either one or two photoabsorption edges simultaneously, paving the way to X-ray dust studies. Fitting one or two edges alone however could lead to degeneracy among different possible solutions and hamper the identification of the dust composition and abundances. In this study we simultaneously fit XAFS from all of the major constituents of interstellar dust (Fe, O, Mg, Si), for the first time, to unveil the exact composition of dust in our Galaxy. The X-ray band provides this possibility only in very special cases, where the source is very bright ($>0.1$\,Crab or $\rm 2.4 \times 10^{-9} \ erg \ cm^{-2} \ s^{-1}$) and the column density is moderate, providing sufficient optical depth and signal-to-noise in the photoabsorption edges of interest. GX 339-4, when in outburst, is one of the unique cases of low-mass X-ray binaries that meet these criteria.

GX 339–4 is a well-studied transient LMXB system comprising a low-mass companion star orbiting a black hole. With an orbital period of 42 hours and located at a distance of approximately 8–12 kpc \citep{Zdziarski}, GX 339–4 is one of the most active transient systems in our Galaxy, having exhibited numerous outbursts over the past decades. Its frequent activity and prominent disk reflection features \citep{Garcia2015, Ludlam2015} make it a key target for studying black hole accretion physics.

Previous studies, such as \cite{Miller2004}, have used broad-band and high-resolution X-ray spectra of GX 339–4 during outburst decline to investigate relativistic effects in the inner accretion disk and ionized outflows. However, in addition to its importance for accretion studies, GX 339–4 is also a valuable probe of the ISM along the line of sight. With a moderate hydrogen column density of $\rm \sim5 \times 10^{21} \ cm^{-2}$ \citep{Kalberla2005}, it provides a rare opportunity to study XAFS in multiple edges—specifically O K ($\sim$ 23 \AA \ or 0.5 keV), Fe L ($\sim$ 17.5 \AA \ or 0.7 keV), Mg K ($\sim$ 9.5 \AA \ or 1.3 keV), and Si K ($\sim$ 6.7 \AA \ or 1.8 keV)—simultaneously. This makes GX 339–4 a unique laboratory for exploring both compact object accretion and the physical properties of the intervening ISM. 

As part of a Chandra Cycle 25 program, we carried out an observational campaign of GX 339–4 during an unusual outburst in fall 2024—its second that year. The source was observed in the soft state at a flux level of approximately 0.3 Crab, ensuring sufficient signal-to-noise ratio for detecting absorption features. In this work we focus on the investigation of the chemical composition of dust grains along the line of sight to the source and to identify absorption features from highly ionized gas. The paper is organised as follows. In Section \ref{datared}, we describe the instrumental setup and data reduction procedures for the Chandra and XMM-Newton datasets. Section \ref{spectralfit} details the spectral fitting methodology used to analyze the gas and dust features, and in Section \ref{disc}, we present a discussion of our results.

\section{Data and reduction}\label{datared}

In Table \ref{tab:obs} we list all the observations used in this paper. To effectively study the narrow absorption features of gas and dust in the soft X-ray band ($\rm <2 keV$), we utilize the high-resolution capabilities of the X-ray spectrometers currently available on board the \xmm and \chandra observatories.
The \chandra observatory is equipped with the High Energy Transmission Grating Spectrometer \cite[HETGS;][]{hetgs}, a high spectral resolution instrument composed of two grating assemblies: the Medium Energy Grating (MEG) and the High Energy Grating (HEG). XMM-Newton is equipped with the Reflection Grating Spectrometer \cite[RGS;][]{denHerder}, which provides high-resolution soft X-ray spectroscopy.

Si K and Mg K edges can be only explored using the HETGS due to its high spectral resolution in this spectral region. This makes Chandra/HETGS the only instrument at the moment of writing this paper, which is suitable to study the Mg K and Si K-edges. Oxygen cannot be studied with Chandra gratings due to contamination buildup on the focal plane detector. Therefore, for the analysis of the oxygen K-edge region, we use RGS data. Both instruments cover the Fe L and Ne K regions. 

\subsection{Chandra/HETGS}
 The new Chandra HETGS observations of 2024 were taken in Timed Exposure mode (TE). The 2-dimensional information provided by TE mode observations (as opposed to 1-dimensional in Continuous Clocking (CC) mode) is necessary to remove distortions around the photoasborption edges, especially in the Si K-edge\footnote{https://cxc.harvard.edu/proposer/POG/html/chap8.html}. In order to suppress photon pile-up in the grating spectra we limited the focal plane configuration to four detectors (S1,S2,S3,S4) and employed a subarray of 350 pixel rows. For more details we refer to the HETGS observer's guide.

The data were processed with the Chandra Interactive Analysis of Observations (CIAO) software, version 4.15, along with the Calibration Database (CALDB) version 4.10.7, as described by \cite{Fruscione2006}. Data reduction and generation of the final grating products, including PHA2 spectra and the RMF and ARF response matrices, were carried out using the {\tt chandra\_repro} script. We utilize both the HEG and MEG data, combining the different spectral orders of each individually with the CIAO tool {\tt combine\_grating\_spectra}. The 120 ks of new observations were divided into seven segments, all conducted within a span of approximately two weeks to ensure the source remained in the soft state, with a flux around 0.3 Crab. 

To cover the softer part of the spectrum—specifically the Fe L and Ne K edge regions—we used archival Chandra data from 2004 during a previous outburst. This choice was made because the instrument's effective area has since declined, making the coverage of these edges with newer data less efficient.

\subsection{XMM-Newton/RGS}
We retrieved the RGS data from the XMM-Newton public archive\footnote{\url{http://nxsa.esac.esa.int/nxsa-web/}}, and processed them using the Science Analysis Software (SAS), version 18. The initial event lists were generated using the \texttt{rgsproc} task. To filter out intervals affected by high particle background, we applied a standard threshold of 0.2 counts/s. Bad pixels were excluded by setting \texttt{keepcool=no} within the \texttt{rgsproc} task.

For bright sources, the RGS spectra can be affected by pile-up, particularly at wavelengths shorter than 19\AA, potentially distorting the spectral shape. To mitigate this, we excluded data below 19\AA \ from our analysis. When handling multiple observations of the same source, we took care to avoid introducing spectral artifacts through improper merging. Observations with consistent spectral shape and flux, or those taken within the same epoch, were combined using the \texttt{rgscombine} command.

\begin{table*}
\caption{\chandra and \xmm observations used in this paper.} 
\label{tab:obs}   
\centering
\scalebox{1}{
\begin{tabular}{c c c c  }     
\hline   
\hline
 obs ID                   &    exp. time (ks) &       date of obs. & Instrument   \\
\hline   
\hline      
\multicolumn{4}{c}{\textbf{Archival XMM-Newton}} \\
\hline                
   0148220201                      &    20                   &  08-03-2003 &  RGS    \\
 0148220301                       &     16                & 20-03-2003     &  RGS  \\
    0654130401                    &    34                 &  28-03-2010    &   RGS        \\
    \hline
    \multicolumn{4}{c}{\textbf{Archival Chandra}} \\
    \hline 
   4569                      &    50                   &  22-08-2004 &  HETGS/MEG    \\
4570                       &     40                & 04-10-2004     &  HETGS/MEG  \\
    4571                    &    36                 &  28-10-2004    &   HETGS/MEG       \\
\hline   
\multicolumn{4}{c}{\textbf{New Chandra}} \\
\hline
    29899                      &    21                   &  26-09-2024 &  HETGS/HEG    \\
    30467                       &     17                & 27-09-2024     &  HETGS/HEG  \\
    30468                   &    15                 &  28-09-2024   &   HETGS/HEG       \\
    30469                  &    15                 &  28-09-2024   &   HETGS/HEG       \\
    30470                   &    15                 &  28-09-2024    &   HETGS/HEG       \\
    30471                   &    20                 &  29-09-2024    &   HETGS/HEG       \\
    30472                   &    18                 &  30-09-2024    &   HETGS/HEG       \\

\hline                
\end{tabular}}
\begin{flushleft}
\footnotesize{   }
\end{flushleft}
\end{table*}

\section{Spectral fitting of the ISM}\label{spectralfit}

\begin{figure*}
    \centering
    \includegraphics[width=1\textwidth]{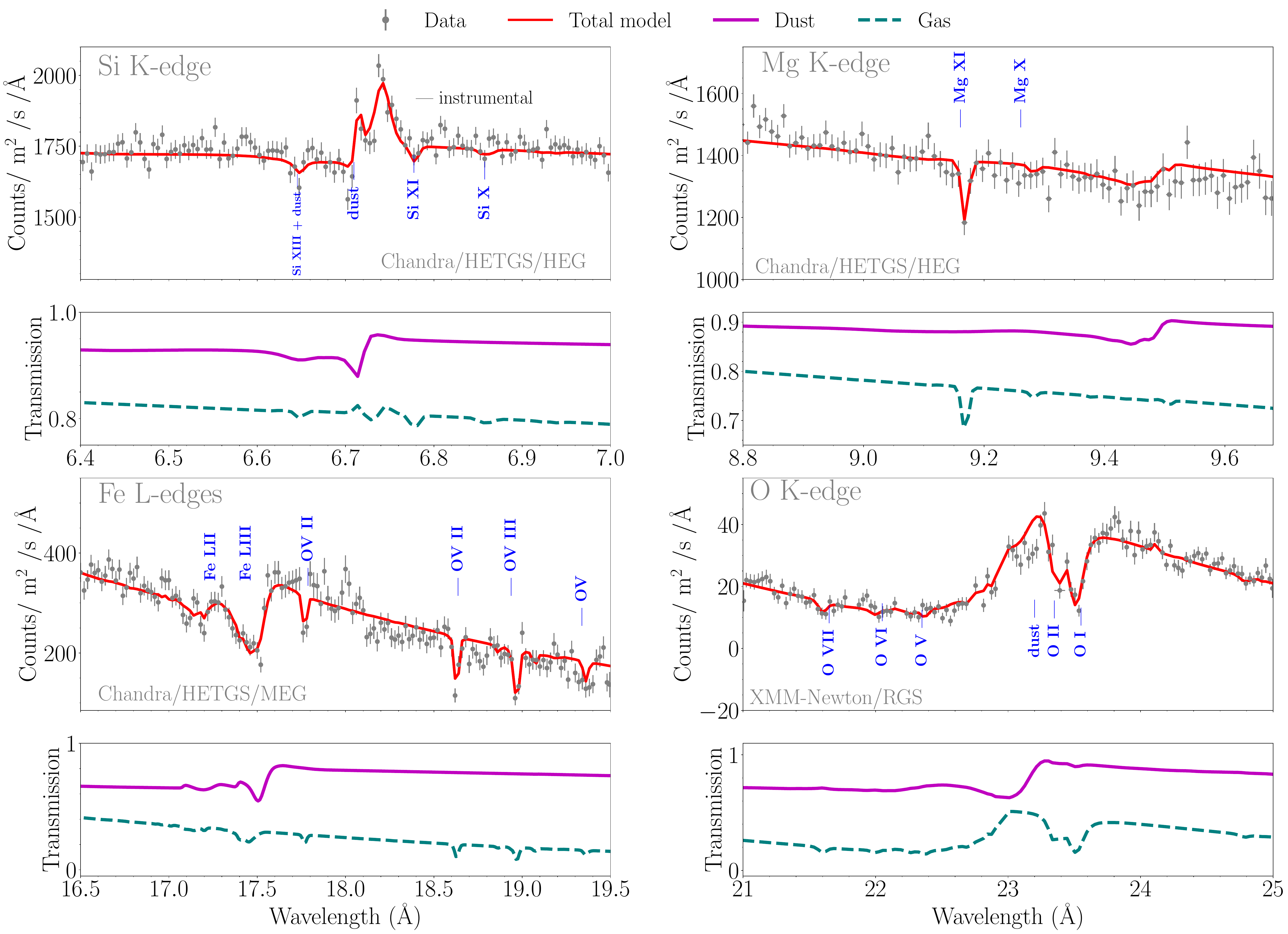}
    \caption{Zoom-in on the Si K, Mg K, Fe L, and O K photoabsorption edges of GX 339-4, and the corresponding transmission of the gas and dust components for each edge. The gas-phase transmission has been shifted on the y axis for presentation purposes. Chandra/HETGS data are used for the Si K and Mg K edges, and XMM-Newton/RGS data for the Fe L and O K edges. The best-fit model includes absorption by neutral and ionized gas, as well as dust. }
    \label{fig:spectra}
\end{figure*}


In this study, we simultaneously fit the photoabsorption edges of O K, Fe L, Si K, Mg K, and Ne K—marking the first comprehensive attempt to study XAFS along the line of sight toward a bright background source (Figure \ref{fig:spectra}). For our spectral modeling, we employ the SPEctral X-ray and UV modelling and analysis software, \textsc{spex}, version 3.08.01\footnote{\url{https://zenodo.org/records/12771915}} \citep{Kaastra1996, Kaastra2018}.

To maximize the spectral resolution available from each instrument, we use the Chandra/HETGS data for the Si K, Mg K, Ne K, and Fe L edges, and the XMM-Newton/RGS data for the O K edge. The spectrum is fitted over the wavelength range from 2 to 35~\AA, a range that is broad enough to fully encompass the edges of interest while excluding regions that could complicate the fits or hinder the XAFS modeling.

Since the Chandra and XMM-Newton observations were taken at different epochs, intrinsic variability in the background source may affect the continuum shape. To account for this, we make use of the \texttt{sectors} functionality in \textsc{spex}, which assigns each dataset to a distinct model. This allows the continuum parameters, which may vary, to be fitted independently for each dataset, while applying a common ISM absorption model across all sectors. Given the narrow spectral range analyzed for each edge, a detailed broadband continuum model is not required. Instead, we adopt a simple phenomenological approach using a power-law component (\texttt{pow}) and a blackbody component (\texttt{bb}) to describe the local continuum around each edge.

We apply a binning factor of 2 to our data, which improves the signal-to-noise ratio while still oversampling the instrument’s spectral resolution, ensuring no loss of accuracy \citep{Kaastra2016}. We use $C$-statistics ($C_{\rm stat}$) to evaluate the goodness of fit \citep{Cash1979, Kaastra2017}. All uncertainties are reported at the 1$\rm \sigma$ (68\%) confidence level. For the abundance values, we adopt proto-Solar units from \citet{Lodders2009}, as the default abundances of \spex. A thorough comparison of the different abundance tables used in ISM studies is presented in detail in \cite{Psaradaki2024}.

\subsection{Modelling the absorption of neutral and ionised gas}

To model the neutral Galactic absorption, we use the \hot model in \spex \citep{dePlaa}. This model computes the ionization balance based on a given temperature and abundance set, and then scales the ionic column densities according to the specified total hydrogen column density. Using this set of column densities, the transmission of the plasma is calculated by multiplying the transmission of the individual ions. At very low temperatures (around 0.001 eV, or approximately 10 K), it effectively reproduces the behavior of neutral gas. The free parameters of this model are the hydrogen column density (\NH) that is calculated from the X-ray absorption, and the temperature of the gas (see Table \ref{tab:ioncol}). For the neutral component, the abundances of O, Fe, Si, and Mg are treated as free parameters, while ensuring they remain within the limits from previous work \citep{Zeegers2019, Rogantini2019, Psaradaki2021}.

To account for absorption from highly ionised plasma that likely originates close to the source, we include two additional \hot components. These components reproduce prominent absorption lines such as \neix and \oviii. 
In Table \ref{tab:ioncol} we summarize the ionic column densities predicted from the \hot models. These ionic column densities are model-dependent and rely on the hydrogen column density of the \hot models; they should therefore be interpreted with caution. Nevertheless, these values remain informative regarding the ions and abundances predicted by each \hot model. 

To model additional absorption features not well captured by the \hot model (e.g., \ov, \nev), we use the \slab model in \spex. This model treats absorption by a slab of optically thin gas, fitting the column densities of individual ions independently, without assuming ionisation balance or a specific physical model. 
 A summary of the detected lines is provided in Table~\ref{tab:ioncol}, and the inclusion of this component improves the $C_{\rm stat}$ value by 780.
 
To accurately reproduce the observed absorption lines, we include two \texttt{slab} components: one at rest and another with a velocity shift ($Z_{v}$) of –540 km/s. While a detailed physical analysis of the origin and properties of this ionised absorber is beyond the scope of this paper, we model these lines to estimate their ionic column densities and to correct residuals in the affected spectral regions. These features likely originate from ionised gas intrinsic to the binary system and material in the environment surrounding the black hole binary GX~339--4. As discussed later, a complete physical model is required to fully understand the nature of these lines.


\begin{table}
\centering
\caption{Column densities of ions relevant to this study, predicted by the \hot\ model or measured using the \slab\ model. For the \hot model, ionic column densities are associated with the corresponding plasma temperature ($kT$) in keV, and hydrogen column density ($N_{\mathrm{H}}$) in $\rm 10^{21} \ cm^{-2}$. For the slab components, column densities are fitted independently and are not associated with ionisation balance.}
\begin{tabular}{lc}
\hline
\hline
Ion & Ionic column density ($10^{15}\ \mathrm{cm}^{-2}$) \\
\hline
\hline
\multicolumn{2}{c}{Hot \#1 ($N_{\mathrm{H}}=4.4\pm0.5$, $kT=10^{-6}$ $fixed$)} \\
\hline
\oi  & $(3.35\pm0.38) \times 10^{3}$   \\
\nei   & $570\pm65$ \\
\fei   & $7.0\pm0.8$ \\
\mgi &  $23.2\pm2.6$\\
\sii & $9\pm1$ \\
\hline
\multicolumn{2}{c}{Hot \#2 ($N_{\mathrm{H}}=0.7\pm0.4$, $kT=(1.65\pm0.05)\times10^{-3}$)} \\
\hline
\oii   & $(3.5\pm2)\times 10^{2}$   \\
\feii    & $7.3\pm4.1$ \\
\feiii     & $13.8\pm7.8$ \\
\mgii    & $9.1\pm5.2$  \\
\mgiii    &  $16.1\pm9.2$ \\
\siii       &  $24.7\pm14.1$   \\
\siiii     &   $0.15\pm0.09$   \\
\hline
\multicolumn{2}{c}{Hot \#3 ($N_{\mathrm{H}}=0.12\pm0.02$, $kT=0.15\pm0.01$)} \\
\hline
\oviii    & $21.4\pm3.5$  \\
\neix    & $15.3\pm2.5$ \\
\mgx     & $0.6\pm0.1$ \\
\mgxi    & $4.2\pm0.7$   \\
\six     & $1.2\pm0.2$  \\
\sixi    & $1.4\pm0.2$   \\
\sixiii  & $0.7\pm0.1$   \\
\hline
\multicolumn{2}{c}{Slab \#1} \\
\hline
\ov      & $<49.7$ \\
\ovi     & $2.4\pm0.2$ \\
\ovii    & $<3.1$ \\
\neii    & $120.23^{+0.01}_{-0.03}$ \\
\neiii   & $32.3^{+2.1}_{-0.1}$ \\
\nevii   & $2.3^{+0.1}_{-0.5}$ \\
\neviii  & $2.1^{+0.01}_{-0.6}$ \\
\hline
\multicolumn{2}{c}{Slab \#2 ($Z_{v} = -540\pm50 \ \rm km \ s^{-1}$)} \\
\hline
\nev     & $<3.5$ \\
\nevi    & $10.30\pm0.01$ \\
\hline
\end{tabular}
\label{tab:ioncol}
\end{table}

\subsection{Fitting the dust absorption}

The presence of dust modifies the shape of the photoabsorption edge, giving a diagnostic for its detection in the X-ray spectrum. To model the dust X-ray absorption fine structures in the spectrum of GX 339-4, we employ the \amol \ model in \spex, which calculates the transmission of a dust component while treating the dust column density as a free parameter. We adopt the dust extinction cross sections from our previous work: detailed descriptions of the dust models are provided in \cite{Zeegers2019} for the Si K-edge, \cite{Rogantini2019} for the Mg K-edge, and \cite{Psaradaki2020, Psaradaki2021} for the O K- and Fe L-edges.\footnote{\href{https://ui.adsabs.harvard.edu/abs/2019yCat..36300143R/abstract}{VizieR Online Data Catalog 1}}\footnote{\href{https://ui.adsabs.harvard.edu/abs/2020yCat..36420208P/abstract}{VizieR Online Data Catalog 2}}

In brief, the dust extinction cross sections were calculated using either Mie \citep{Mie} or Anomalous Diffraction Theory \citep[ADT;][]{van_de_Hulst}, assuming a Mathis-Rumpl-Nordsieck (MRN) dust size distribution \citep{mathis}. This distribution follows a power-law of the form $dn/da \propto a^{-3.5}$, where $a$ represents the grain size, with a lower cutoff at $\rm 0.005 \mu m$ and an upper limit of $\rm 0.25 \mu m$. In Table \ref{tab:samples} we summarize the different types of dust minerals used in this study. The samples 2,5,11 are natural and 8,10,12 are commercial products. Samples 1,3,4,6,7,9 are instead synthesized in the laboratories at the Astrophysikalisches Institut, Universitats-Stenwarte (AIU), and Osaka University. We adopted the metallic iron presented by \citet{Kortright} and \citet{Lee2010}, with a shift in energy according to the work of \citet{Fink}. The extinction cross section data for the majority of the models are publicly available in online catalogues (see, e.g., \citealt{Rogantini2019, Psaradaki2020}).

The \amol model can fit up to four different dust compounds simultaneously at a given fitting run. We test all the possible combinations of four dust species among the 13 samples assuming that the interstellar dust mixture can be described with at most 4 components, previously discussed in \cite{Costantini2012}. The number of different combinations is given by the equation $C_{e,c}=e!/c!(e-c)!$, where $e$ is the number of the available edge profiles and $c$ the combination class. This gives us 715 dust mixtures to choose from when fitting the spectra for each source. 

\begin{figure*}
    \centering
    \includegraphics[width=\textwidth]{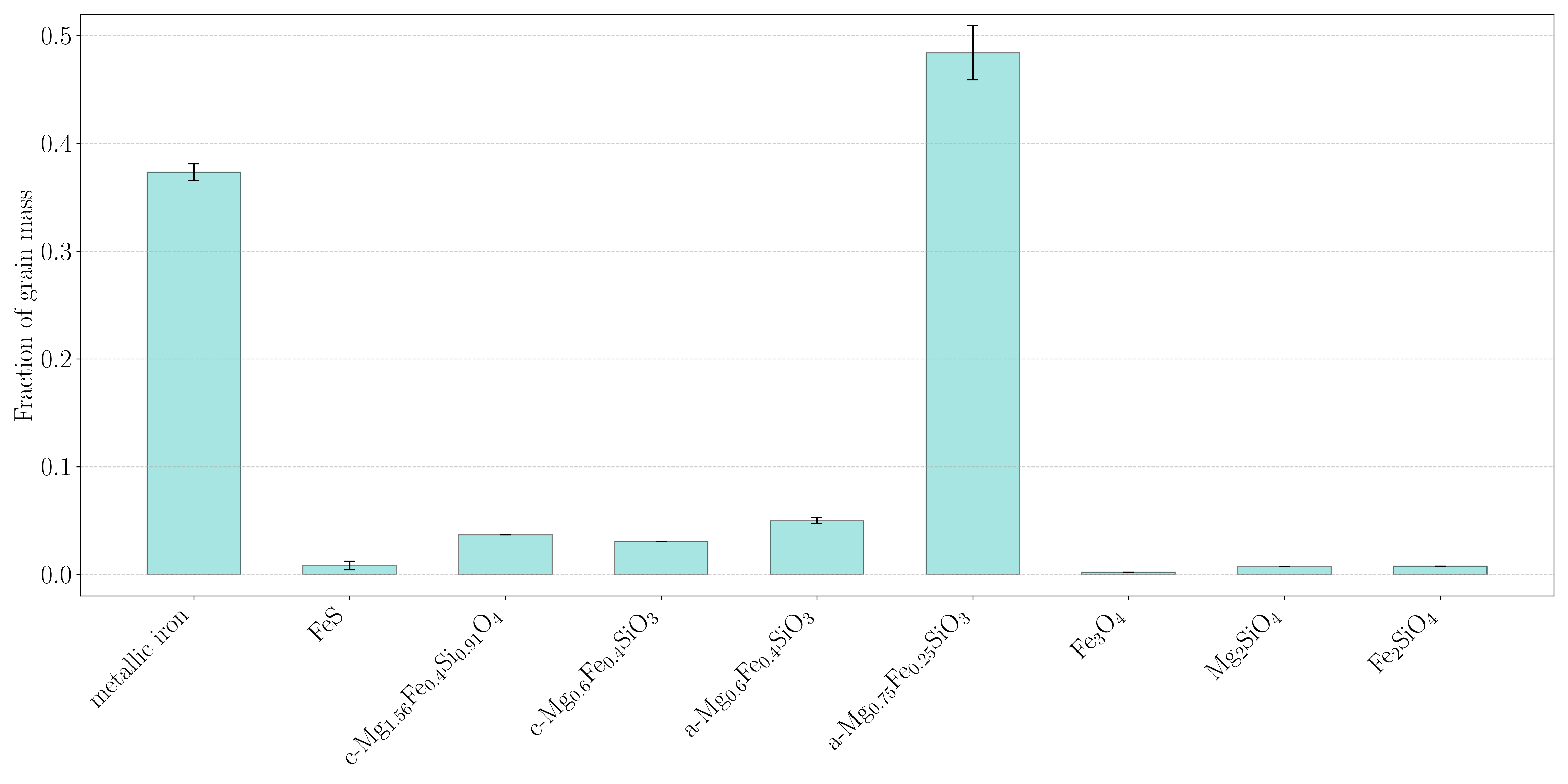}
    \caption{Relative fraction of column density for each dust compound. The fractions have been calculated considering models with $\Delta \mathrm{AIC} < 4$. The symbol $a$ denotes amorphous compounds, while $c$ refers to crystalline ones. The error bars indicate the minimum and maximum percentage of each compound within the $\Delta \mathrm{AIC} < 4$ selected models.}
    \label{fig:barplot}
\end{figure*}

We evaluate the best-fitting dust mixture using the Akaike Information Criterion (AIC; \citealt{Akaike}), a widely used and efficient tool for model selection, as described in \cite{Rogantini2019}. AIC allows for rapid comparison of models that are statistically comparable to the best fit. Unlike traditional likelihood-based methods that assume a fixed model structure and estimate its parameters accordingly, AIC generalizes the approach by also allowing the model dimension to vary. \citep{Feigelson}.
Importantly, AIC treats all models symmetrically and does not require that one of the candidate models be the “true” model. This makes it suitable for comparing both nested and non-nested models. In our case, the candidate models are non-nested and share the same number of free parameters.
In practice, we compare the C-statistic values of each fit to that of the best-fitting model. The AIC is then calculated as:

\begin{align*}
\rm AIC=2\textit{k}-2ln(\mathcal{L}_{max})
\end{align*}

\noindent where $k$ is the number of fitted parameters of the model and $\mathcal{L}_{max}$ is the maximum likelihood value. The C-statistic is related to the maximum likelihood through the relation $\rm Cstat = -2\ln(\mathcal{L})$ \citep{Cash1979}. While we are not interested in the absolute AIC value, we focus on the AIC difference ($\rm \Delta AIC$) for each candidate model relative to the model with the lowest AIC. Models with $\rm \Delta AIC < 4$ are considered statistically equivalent and provide similarly good fits to the data, whereas models with $\rm \Delta AIC > 10$ can be ruled out (\citealt{Burham}). 
In Figure \ref{fig:barplot} we show the results of the relative fraction of column density for each dust compound. The dust fraction comes from all the model selected via their AIC values, using models with $\Delta \mathrm{AIC} < 4$. The symbol $a$ represents amorphous compounds, while $c$ denotes crystalline compounds. The error bars show the minimum and maximum percentages of each compound across the models selected with $\Delta \mathrm{AIC} < 4$. Finally, Figure \ref{fig:spectra} provides a zoom-in of all photoabsorption edges and the corresponding best-fit model with $C_{\rm stat}/d.o.f.=4497/3202$, where $d.o.f$. is the degrees of freedom. The dust component has improved the fit by $\Delta C_{ \rm stat}=880$.

\begin{table}
\caption{List of dust samples.}
\begin{tabular}{cccc}
  \hline
  \hline
 \# & Compound & Chemical Formula & Form\\
  \hline
  1 &  Olivine & $ \rm MgFeSiO_{4}$ &  amorphous \\
  2 &  Olivine & $ \rm Mg_{1.56}Fe_{0.4}Si_{0.91}O_{4}$ &  crystalline \\
 3 &  Pyroxene & $ \rm Mg_{0.6}Fe_{0.4}SiO_{3}$ &  amorphous \\
 4 &  Pyroxene &  $ \rm Mg_{0.6}Fe_{0.4}SiO_{3}$ &  crystalline \\
 5 & Enstatite & $ \rm MgSiO_{3}$ &  crystalline \\
  6 & Enstatite & $ \rm MgSiO_{3}$&  amorphous \\
 7 &  Fayalite & $ \rm Fe_{2}SiO_{4}$ &  crystalline \\
 8 &  Forsterite & $ \rm Mg_{2}SiO_{4}$ & crystalline \\
 9 & Pyroxene & $ \rm Mg_{0.75}Fe_{0.25}SiO_{3}$ &  amorphous \\
 10 & Magnetite & $ \rm Fe_{3}O_{4}$ &  crystalline \\
 11 &  Troilite & $ \rm FeS$ &  crystalline\\
 12 &  Pyrit Peru & $ \rm FeS_{2}$ &  crystalline\\
 13 &  Metallic iron & $ \rm Fe$ &  -         \\
\hline
\hline
\end{tabular}
\label{tab:samples}
\end{table}

\section{Discussion}\label{disc}

\subsection{Dust mineralogy towards GX 339-4}

Our analysis indicates that Mg-rich amorphous pyroxene ($\rm Mg_{0.75}Fe_{0.25}SiO_{3}$) and metallic iron are the dominant dust components. Table \ref{tab:dust} presents the column densities of the best-fit dust compounds, while Figure \ref{fig:barplot} illustrates the relative contribution of each compound to the overall dust composition. This plot includes models with $\Delta AIC <  4$ to ensure statistical robustness. We find that Mg-rich amorphous pyroxene contributes approximately 50\% of the dust column, while metallic iron accounts for about 40\%. The remaining dust is primarily composed of other silicates, such as $\rm Mg_{0.6}Fe_{0.5}SiO_{3}$, which is also Mg-rich and Fe-poor. We do not find significant amounts of iron sulfides (FeS) or magnetite ($\rm Fe_{3}O_{4}$) along this sightline.

\begin{table}
\caption{Column density of best fit dust combination, in units of $\rm 10^{16} \ cm^{-2}$.} 
\label{tab:dust}   
\centering
\begin{tabular}{c c }     
\hline   \hline
Compound         &       $N_{dust} $        \\
\hline \hline
Metallic iron   & $\rm 10.6\pm0.4$       \\
c-$\rm Mg_{2}SiO_{4}$ & $0.4\pm0.2$   \\
c-$\rm Mg_{0.6}Fe_{0.4}SiO_{3}$ & $4.4^{+0.7}_{-4.2}$  \\
a-$\rm Mg_{0.75}Fe_{0.25}SiO_{3}$ & $13.1^{+0.6}_{-0.1}$ \\
\hline \hline       
\end{tabular}
\end{table}

This result is broadly consistent with our previous work using the same dust models computed from the laboratory data \citep{Psaradaki2020}. In \cite{Psaradaki2022} we analyzed a sample of five sightlines along the Galactic plane and found that $\rm Mg_{0.75}Fe_{0.25}SiO_{3}$ is the primary dust reservoir in the diffuse ISM. In that study, the contribution from metallic iron was lower—averaging around 15\% across all sightlines. However, it is important to note that GX 339–4 is significantly brighter than the sources in the earlier sample, and the data around the Fe L edges have a higher signal-to-noise ratio. Additionally, we find that the Fe abundance toward GX 339–4 is close to Solar, compared to the previous study where we found under-abundance of iron, raising the possibility that the previously ``hidden'' reservoir of Fe is, in this case, primarily in metallic form.

Our results on the silicate composition of interstellar dust are broadly consistent with previous infrared studies. For example, \citet{Min2007} found that interstellar silicates are magnesium-rich, with compositions between pyroxene and olivine, which aligns with the presence of Mg-rich silicates in cometary and circumstellar environments. Similar conclusions were drawn by other studies modeling the 10 $\rm \mu$m silicate feature using mixtures of Mg-rich olivines and pyroxenes (\citealt{Molster2002}, \citealt{Chiar2006}, \citealt{Fogerty2016}). 

Since the amorphous olivine in our sample is limited to Mg/Fe = 1, we cannot exclude the possibility that the inclusion of Fe-poor amorphous olivines would alter the relative pyroxene/olivine balance in our results. Another factor that could potentially influence the results is the dust grain size distribution, which can affect the shape of the dust scattering peak. This effect is particularly pronounced at the O K and Fe L edges. We also note that the dust scattering properties also depend on grain porosity, and allowing for non-negligible porosities could, in principle, affect the inferred composition. Exploring the combined effects of porosity and grain size is beyond the scope of the present work and will be examined in a follow-up study.

\subsection{Abundances and depletions of O, Fe, Si, Mg}

\begin{table*}
\caption{Abundances and depletions of elements.} 
\label{tab:abund}  
\centering
\begin{tabular}{c c c c c c c}    
\hline   \hline
Element          &       $ N_{gas} $              &   $ N_{dust}$                &  $A_{gas} $              &	$A_{dust} $   & $A/A_{\odot}$ & $\delta_{X}$  \\
        &  $ \rm 10^{15} \ cm^{-2}$  &  $ \rm 10^{17} \ cm^{-2} $   &  $ \rm 10^{-6} \ H^{-1} $ & $\rm 10^{-5} \ H^{-1} $ &  & \\
\hline \hline
Oxygen  & $(3.35\pm0.38) \times 10^{3}$  & $5.3\pm0.3$ & $757\pm105$ & $11.9\pm0.7$ & $1.4\pm0.1$ & $0.14\pm0.02 $      \\
Iron    & $7.0\pm0.8$ & $1.6\pm 0.1$ & $1.6\pm1.2$ & $3.6\pm0.1$ & $1.2\pm0.1$ & $0.96\pm0.13$  \\
Silicon &  $9\pm1$ & $1.8\pm0.1$ & $2.1\pm0.2$ & $4.0\pm0.2$  & $1.1\pm0.1$ & $0.95 \pm 0.12$    \\
Magnesium & $23.2\pm2.6$ & $1.25\pm0.06$ & $5.3\pm0.7$ & $2.9\pm0.1$ & $0.85\pm0.09$ & $0.84\pm0.10$\\
Neon      &    $570\pm65$   &     -          &   $(1.6\pm0.2)\times10^{2}$            &      -      &     $1.27\pm0.14$          &      -         \\
\hline \hline       
\end{tabular}
\begin{flushleft}
\footnotesize{ $N_{ \rm gas}$ and $N_{ \rm dust}$ indicate the total column density of gas and dust respectively, in $\rm cm^{-2}$. $ A_{\rm gas}$ and $A_{\rm dust}$ are the abundances of the elements in gas and dust respectively (with respect to hydrogen). $ A/A_\odot$ is the total abundance (gas+dust) ratio in proto-Solar abundance units, and $\delta_{X}$ is the depletion of this element into solids. The last row shows the values of (neutral) neon, which, as a noble gas, exists only in the gas phase.} 
\end{flushleft}
\end{table*}

Table \ref{tab:abund} summarizes the calculated elemental abundances and depletions for the key dust-forming elements studied in this work: Mg, Fe, O, and Si. These values are derived from the best-fit spectral model. The first two columns list the column densities in the gas phase ($N_{\mathrm{gas}}$) and in dust ($N_{\mathrm{dust}}$), respectively. The dust column densities can also be inferred from Table \ref{tab:dust} by accounting for the number of atoms in each molecular compound. For instance, if a dust species contains three oxygen atoms, its column density in Table \ref{tab:dust} should be multiplied by three to yield the total oxygen contribution to $N_{\mathrm{dust}}$ as shown in Table \ref{tab:abund}.
Columns 3 and 4 present the elemental abundances in the neutral gas and in dust, each normalized to hydrogen. Column 5 shows the total abundance (gas + dust) relative to the proto-Solar values of \citet{Lodders2009}. Finally, Column 6 gives the depletion factor for each element, defined as the fraction of the total elemental abundance that is incorporated into dust grains. 

\textbf{Oxygen:} We find that oxygen is depleted by approximately 14\% into dust grains and appears to be overabundant relative to the proto-Solar reference. This result aligns broadly with our previous studies of various Galactic sightlines \citep{Psaradaki2020, Psaradaki2022}, as well as with earlier findings by \citet{Costantini2012} and \citet{Pinto2013}. These studies likewise reported oxygen overabundance—though with varying values across different sightlines—and evidence for mild oxygen depletion into dust. In \citet{Psaradaki2024}, we discussed the range of Solar oxygen abundance values reported in the literature, highlighting differences from the \citet{Lodders2009} standard adopted in \spex. Accounting for uncertainties, the total (gas + dust) oxygen abundance along the Cygnus X-2 sightline slightly exceeds the Solar value across all reference scales. Taken together, these results suggest that the commonly adopted Solar oxygen abundance may represent a lower limit, and that the actual oxygen abundance—particularly along the Galactic plane where our X-ray sources reside—could be higher than the standard Solar values used for this highly abundant ISM element. Alternatively, an underestimation of the hydrogen column density could lead to a slightly higher inferred oxygen abundance.

\citet{Gatuzz2016} studied the O K-edge in a sample of Galactic sources and reported sub-Solar oxygen abundances along most sightlines. However, their analysis did not include a dust component in the modeling. Since a significant fraction of oxygen is expected to be locked in dust—especially in silicate grains, which typically contain three or four oxygen atoms—excluding the dust contribution likely leads to an underestimation of the total oxygen abundance. Our results further support the importance of accounting for both gas and dust phases when assessing elemental abundances in the interstellar medium, particularly for elements like oxygen that are major constituents of dust.

\textbf{Iron:} We find that the iron abundance toward GX 339–4 is consistent with, or slightly exceeds, the Solar value, with approximately 96\% of the iron depleted into dust grains. This level of depletion is consistent with the findings of \citet{Jenkins2009}, who analyzed over 200 sightlines within 3 kpc of the Milky Way and reported that, at most, only 10\% of iron resides in the gas phase. Such strong depletion of iron into dust has also been confirmed by our previous survey studies using Chandra/HETGS \citep{Psaradaki2022, Psaradaki2024}, as well as by other X-ray investigations \citep[e.g.,][]{Pinto2013, Costantini2012}. 

In this study, iron is found to be slightly overabundant relative to the Solar reference. This is consistent with earlier work on GX 339-4 by \cite{Corrales2024} who studied the Fe L using XMM-Newton/RGS data. Our prior analysis of five diffuse sightlines along the Galactic plane \cite{Psaradaki2022} indicated that iron was marginally under-abundant. A key difference in the present work is the higher column density of metallic iron, as well as the inclusion of four absorption edges in the fit, compared to only two in the earlier study. The GX 339–4 data benefit from a much higher signal-to noise ratio, owing to the brightness of the source, which likely yields more robust constraints on the Fe L spectral region. This suggests that metallic iron could possibly
represent a previously underestimated reservoir of interstellar iron.

\textbf{Magnesium:} We find that magnesium is predominantly depleted into dust, with approximately 84\% of it residing in the solid phase. \cite{Rogantini2020} analyzed several lines of sight along the Galactic plane, simultaneously examining the Si K and Mg K edges using the same dust models that are employed in this study. Our measured depletion value is slightly lower than the average reported in their sample. This discrepancy may be attributed to differences in the ISM sightline. In particular, GX 339-4 exhibits a lower column density than the sources studied by \cite{Rogantini2020}, and therefore probes more diffuse ISM regions. 
The magnesium abundance appears to be slightly sub-Solar, suggesting that some magnesium could possibly be intrinsically missing. The Mg K-edge is notably weaker in this spectrum compared to other edges like Fe L and O K, which tend to dominate the dust model fits.

\cite{Miller2004} studied the Chandra/HETGS spectrum of GX 339-4, and as part of their analysis, they examined individual ISM photoelectric absorption edges using local models consisting of power-law continua with simple edge functions. They identified neutral (atomic) O, Fe L3 and L2, Ne, Mg, and Si edges. Although their approach did not involve detailed ISM modeling, they derived elemental abundances of 0.75–1.0 relative to Solar, for all elements including Mg, consistent with the Mg abundance we report here.

\textbf{Silicon:} We find that silicon is predominantly depleted into dust grains, with approximately 95\% of it locked in dust. The overall abundance of silicon appears to be close to Solar. This result aligns with previous studies of the Si K-edge, such as those by \cite{Zeegers2019} and \cite{Rogantini2020}. Silicon is known to be one of the elements most heavily depleted into dust, as also demonstrated by \cite{Jenkins2009}. This is consistent with geological observations: silicate minerals are widespread on Earth, the Moon, Mars, and in meteorites. Crystalline silicates, in particular, exhibit strong absorption features near 10 $\mu$m. There is clear evidence that silicate materials are abundant in the interstellar medium \citep{Drainebook}, and the depth of the silicate absorption feature in the infrared indicates that much of the interstellar silicon is incorporated into amorphous silicate grains \citep[e.g.][]{Kemper2004}, as we confirm in this study.

\subsection{What are the primary reservoirs of iron in the dust phase?}

Understanding the exact reservoirs of iron in the diffuse ISM remains an active area of study. It has been hypothesized that most iron depletion occurs within the ISM itself, rather than in standard condensation environments such as AGB stars. In this scenario, iron is injected into the ISM primarily in gaseous form, leaving open the question of what phase or compound it is incorporated into \cite[e.g.][]{Dwek2016, Zhukovska2018}.

Iron exhibits two types of absorption edges in the X-ray spectrum: the Fe K-edge at 7.112 keV and three Fe L-edges, located at 0.846, 0.721, and 0.708 keV (LI, LII, LIII). Therefore, X-ray observations offer a uniquely powerful method for studying iron in the interstellar medium, given the distinctive absorption edges exhibited in the X-ray absorption spectrum.
In this work we find that approximately 65\% of iron is in metallic iron, 32\% in silicates and 3\% in other compounds such as FeS. This result comes from the calculation of the mass density of iron atoms that reside in each chemical compound of the best fit.

\cite{Westphal} suggested that Fe is evenly partitioned between sulfides and metal, a distribution consistent with Fe and S being confined to these phases in a high-temperature condensation sequence. However, spectral fitting indicates that the ISM does not contain a significant fraction of Fe sulfides.
In \citet{Psaradaki2022}, we analyzed five diffuse sightlines using X-ray spectroscopy and found that approximately 40\% of the dust-phase iron is in metallic form, with the remaining 60\% incorporated into silicates. These values represent an average across all five sources examined. For GX 339–4, we find a higher fraction of metallic iron compared to the earlier study. Notably, in the previous work, we reported an underabundance of iron relative to Solar, whereas in the present analysis we recover the full expected iron content. This discrepancy may suggest that the sightline toward GX 339–4 is intrinsically different, or alternatively, that the higher signal-to-noise ratio near the Fe L-edge, due to the brightness of the source, allows for a more reliable modeling of metallic iron.

Finally, \citet{Draine2001} estimated that less than 10\% of solid-phase iron resides in metallic form, primarily due to the expectation that ferromagnetic iron would produce distinctive thermal magnetic dipole emission \citep{Draine2013}. However, \citet{Hensley_2023} discuss that earlier models may have overestimated this emission, potentially allowing for a greater fraction of metallic iron in dust without conflicting with polarization constraints. The significance of metallic iron as a major dust component therefore remains an open question.

Further laboratory measurements of metallic iron are especially important, as it remains the only dust compound in our analysis that has not yet been directly characterized in our previous experimental campaigns. Earlier studies \citep{Psaradaki2020, Corrales2024} have pointed out inconsistencies in the energy calibration of metallic iron features used in X-ray analyses and emphasized the need for updated laboratory data. Improved laboratory benchmarks would help refine the modeling of metallic iron and reduce associated uncertainties in X-ray absorption studies. 

\subsection{The Ne K-edge and absorption lines from highly ionised plasma}

\begin{figure}
    \centering
    \includegraphics[width=0.48\textwidth]{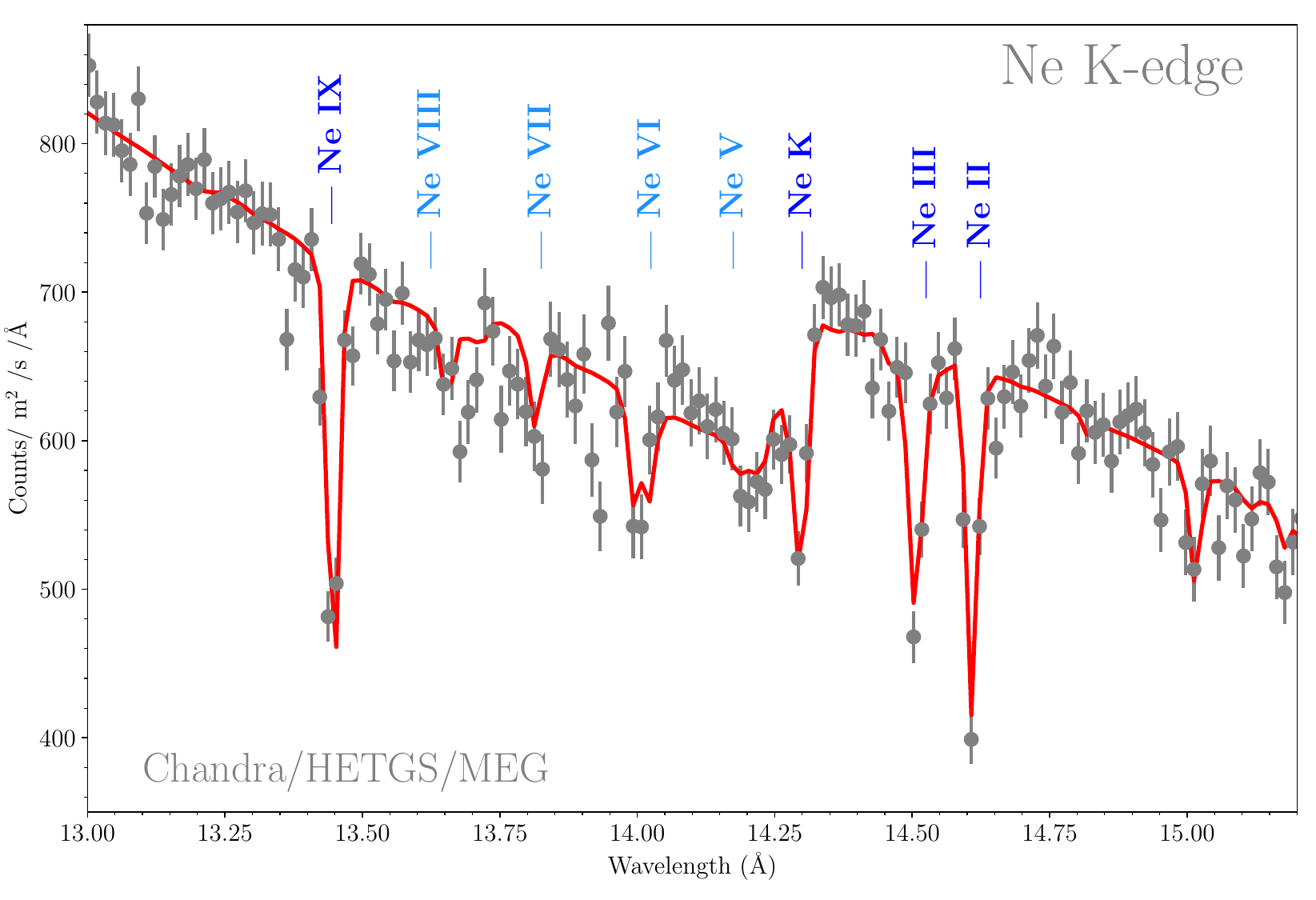}
    \caption{Zoom-in on the Ne K photoabsorption edge of GX 339-4. Significant absorption by \nei, \neii, \neiii, and \neix is detected. Residuals in the 13.5–14.25\AA \ range may indicate the presence of additional ionized neon species; however, we exercise caution in this interpretation as a full photoionization model is required to confirm this.}
    \label{fig:neon}
\end{figure}

In addition to the four absorption edges of primary interest—Mg K, O K, Si K, and Fe L—which are relevant for studying interstellar dust, we also fit the Ne K edge simultaneously in our spectral modeling (Figure \ref{fig:neon}). Neon is a noble gas and, as such, is not expected to be incorporated into dust grains. In our spectrum, we clearly detect the Ne K edge, along with strong and well-resolved absorption lines from Ne II, Ne III, and Ne XIII. The column densities derived for these ions are very well constrained. 

These lines have been previously detected in the X-ray spectrum of GX 339-4. \cite{juett2006} found that \neii and \neiii absorption is consistent with an origin in the warm ionised medium (WIM), rather than in the neutral ISM. Moreover, \cite{Miller2004} report strong Ne II and Ne III absorption lines in the Chandra high-resolution spectrum of the source and find that their widths exceed those expected from thermal broadening in the coronal ISM, favoring an origin in a local, intrinsic warm absorber with moderate turbulent velocity broadening. In this work, we do not attempt a detailed analysis of the line widths, as our primary focus is on the dust component. However, considering the above, a plausible interpretation is that the observed Ne II and Ne III absorption arises from a combination of ionized regions in the ISM and additional ionized material local to the black hole. Neon X-ray absorption in the local ISM has also been studied in detailed from \cite{Gatuzz2016}.

We compared the Ne with the O abundance derived from this study. We followed the same methodology as in our previous analysis of Cygnus X-2 \citep[][Table 5]{Psaradaki2024}. We calculated the log(O/Ne) ratio as the sum of the predicted \oi abundance that we obtain from the fit, and the solid phase oxygen. The Ne abundance is determined through the fit of the Ne K-edge, and in particular it represents the summed abundance of \nei, \neii and \neiii. We find that log(O/Ne)=0.67 which is consistent with the reference abundance tables in Table 5 of \citet[][]{Psaradaki2024}.

In addition, we observe a set of residuals in the 13.5–14.25\AA \ range. To investigate their origin, we tested various lines and fitted them using slab components for Ne V, Ne VI, Ne VII, and Ne VIII. In order to achieve a good fit for Ne V and Ne VI, a velocity shift was required (Table \ref{tab:ioncol}), which may indicate the presence of a distinct warm absorber. However, other species such as \ovi\ and \ovii\ do not require a velocity shift, suggesting that the observed shifts in these neon lines may instead arise from limitations in the atomic database or calibration uncertainties for those specific transitions. 

Similar features have been reported in the Chandra/HETGS spectrum of Cygnus X-2 by \citet{Schulz2002}, who identified them as high-ionisation iron lines (Fe XIX and Fe XX), suggesting a potentially different physical origin. In our analysis, however, these iron lines, when modeled using the available atomic data in \spex, do not provide a good fit to the features observed in the 13.5–14.25\AA \ range. Moreover, \cite{Gatuzz2015} studied ISM absorption features for a number of Galactic X-ray sources, and identified some of the neon lines to be associated with higher resonances of \neii and \neiii. The origin of these residuals remains uncertain, and a detailed investigation involving more complete physical modeling is left for future work, as it lies beyond the scope of this paper. Importantly, these neon features are well separated from the spectral regions associated with dust absorption, and their presence does not affect our results.

Around the Si and Mg K edges, we also detect absorption lines from ions of higher ionisation stages. In particular, we identify transitions from \mgx, \mgxi, \six, \sixi, and \sixii. Ionised absorption near the Si K-edge has also been found in previous ISM studies \cite[e.g.][]{Schulz2016, Gatuzz2020, Yang2022}. These ions suggest the presence of hot, ionised gas along the line of sight. In addition, the spectrum shows a rich set of oxygen transitions ranging from \oii to \oviii. It is likely that the lower-ionisation species (e.g., \oii- \oiv) originate from the ISM, while the higher-ionisation lines, including those of Si, Mg, and \ovii– \oviii, are more consistent with a local absorber, possibly associated with circumbinary material or a disk wind. Some of these high-ionisation lines have ionisation potentials comparable to those of the neon ions we discussed earlier, such as \nev–\neix, which were also attributed to a hotter, more local component. We note, however, that the exact correspondence in ionisation potential between species like \nevi and, for instance, \mgix or \ovi, depends on detailed photoionisation modeling such as that of \cite{Rogantini2021}, and we defer a more quantitative comparison to upcoming future work.

\subsection{Advantages of fitting 4 photosabsorption edges simultaneously}

In \cite{Psaradaki2022}, we demonstrated that a simultaneous fit of the O K and Fe L edges significantly improves our ability to constrain the composition and depletion of interstellar dust, compared to fitting a single edge alone. In that study, although we employed a physically-motivated dust model and achieved a clear separation between gas and dust components in the O K-edge, our ability to distinguish between specific dust species remained limited. The inclusion of the Fe L-edges provided additional information, as the iron content in silicates and metallic forms introduces distinctive spectral features, which help break the degeneracy among chemically plausible dust mixtures. Consequently, the number of acceptable fits (as evaluated via AIC) was reduced, and the resulting depletion and abundance estimates—particularly for oxygen—became more robust. For instance, we revised the oxygen depletion toward Cyg X-2 from 7\% (where only olivine was assumed) to 18 ± 2\% in the current work, based on a broader mineralogical basis.

In this study, we extend this approach further by incorporating simultaneous fits of the Mg K and Si K edges, in addition to the O K and Fe L edges. This advancement allows us, for the first time, to directly constrain the Mg, Si, Fe, O abundances and depletions in interstellar dust from X-ray spectroscopy simulteneously. This approach of fitting all four edges together gives us a clearer view of what interstellar dust is made of. It allows us to measure how much oxygen, iron, magnesium, and silicon is locked up in dust more consistently.

\subsection{Instrumental features around the Si K-edge}

We identify two emission-like features near the silicon K-edge, located at 6.71 and 6.74 \AA. These features have previously been observed in HEG spectra of low-\NH sources by \citet{Rogantini2020}. In our analysis of GX 339–4, we detect these features at a significance above 5$\sigma$. They are considered instrumental in origin, as discussed by \citet{Rogantini2020} and \cite{Yang2022} and further detailed in the HETGS calibration documentation.\footnote{\url{https://space.mit.edu/CXC/calib/hetg_user.html}} \cite{Rogantini2020} attribute these features to instrumental effects based on their observation that only the +1 MEG arm— which does not intersect the front-illuminated CCD in the silicon energy range—lacks the emission peak. This suggests that the pronounced calibration feature associated with the silicon edge in the effective area of the -1 MEG and $\pm$1 HEG arms may contribute to the appearance of this spike. In the GX 339–4 spectrum, these features appear particularly prominent, indicating they are unlikely to be caused by X-ray dust scattering effects, as previously discussed in \citet{Schulz2016}. This provides additional strong evidence supporting their instrumental origin.

\section{Conclusions}

In this work, we used 120 ks of new Chandra/HETGS observations to study interstellar absorption toward the black hole binary GX339–4. The X-ray source is a well-studied low-mass X-ray binary consisting of a black hole accreting from a low-mass companion star, and it also serves as an excellent probe of the ISM. With a moderate hydrogen column density of $\rm \sim 5\times 10^{21} \ cm^{-2}$ and strong X-ray flux, it provides a rare opportunity to simultaneously study X-ray absorption fine structures (XAFS) in multiple edges—specifically O K, Fe L, Mg K, and Si K. These new Chandra/HETGS observations, taken during an unusual outburst in fall 2024, provide a valuable dataset for investigating the properties of cosmic dust in the diffuse ISM. We combined them with archival XMM-Newton/RGS spectra to study the intervening ISM along the line of sight to GX~339–4. The main results of our analysis are summarized below.

\begin{itemize}

\item Our analysis indicates that the dust grain composition along this diffuse Galactic sightline is dominated by Mg-rich amorphous pyroxene ($\rm Mg_{0.75}Fe_{0.25}SiO_{3}$) and metallic iron. Mg-rich pyroxene accounts for roughly 50\% of the total dust column, while metallic iron contributes about 40\%. The remaining fraction consists primarily of other silicate species, such as $\rm Mg_{0.6}Fe_{0.4}SiO_{3}$, which are similarly rich in magnesium and less abundant in iron. We find no significant evidence for iron sulfides (FeS) or magnetite ($\rm Fe_{3}O_{4}$) along the line of sight.

\item In this work, we find that roughly 65\% of the total iron is locked in metallic iron, 32\% in silicate grains, and the remaining 3\% in other compounds such as FeS. These estimates are based on calculations of the mass density of iron atoms associated with each compound in the best-fitting dust model.

\item From a simultaneous spectral fit of the Fe, O, Si, and Mg photoabsorption edges, we constrained both the elemental abundances and depletions. We confirm that Fe, Si, and Mg are predominantly locked in dust, with depletion levels of approximately 85–95\%, while oxygen shows mild depletion, with up to 15\% in the dust phase. The elemental abundances are found to be Solar or slightly above Solar, compared to the reference values of \citet{Lodders2009}.

\item We confirm the presence of absorption features from highly ionised gas in the spectrum of GX~339–4, including lines such as \ovii\ and \neix. This ionised material could potentially be intrinsic to the source and originates in the environment surrounding the black hole binary.

\end{itemize}

Detailed simultaneous fitting of four photoabsorption edges with gas and dust models is feasible only for a limited number of X-ray sources that combine moderate column densities with high flux. This gives valuable insights into the dust grain composition and the abundances of elements predominantly locked in dust. With XRISM, we now have access for the first time to the Fe K-edge spectral region, enabling future inclusion of this feature in simultaneous multi-edge fitting. Furthermore, the upcoming launch of NewAthena will allow detailed investigations of Fe-bearing dust grains in the dense interstellar medium.

\begin{acknowledgments}

We would like to thank the referee for the valuable comments that helped to improve this paper. IP acknowledges support through the European Space Agency (ESA) Research Fellowship in Space Science.
Support for this work was provided by the National Aeronautics and Space Administration through Chandra Award Number GO5-26099X issued by the Chandra X-ray Observatory Center, which is operated by the Smithsonian Astrophysical Observatory for and on behalf of the National Aeronautics Space Administration under contract NAS8-03060.
 Additional support was provided by NASA through the Smithsonian Astrophysical Observatory (SAO) contract SV3-73016 to MIT for Support of the Chandra X-Ray Center (CXC) and
Science Instruments. CXC is operated by SAO for and
on behalf of NASA under contract NAS8-03060.

\end{acknowledgments}

\newpage

\bibliography{reference}{}
\bibliographystyle{aasjournal}


\end{document}